# Generic Forward Curve Dynamics for Commodity Derivatives

David Xiao

## Abstract

This article presents a generic framework for modeling the dynamics of forward curves in commodity market as commodity derivatives are typically traded by futures or forwards. We have theoretically demonstrated that commodity prices are driven by multiple components. As such, the model can better capture the forward price and volatility dynamics. Empirical study shows that the model prices are very close to the market prices, indicating prima facie that the model performs quite well.

## 1. Introduction

Unlike financial instruments in other asset classes, such as Equity, Interest Rate, and FX, commodity forwards and futures are often modeled without a drift. This is because a very large amount of historical data would be required to calibrate any sort of reasonable drift. Commodities rise and fall over time. They are generally difficult to associate a drift to such motion.

Some commodities, such as natural gas, exhibit seasonality in prices. For example, in the winter months, the higher demand for natural gas raises the prices. Therefore, commodity prices are simulated with a seasonal average and a random component. Considering that the linear drift is zero, the seasonal average is constant over time.

The random component of the commodity forward and future prices is generally broken down into major contributors or factors. These are known as the principal components. The principal components for the same commodity are generally independent – that is, we assume that there exists in the world N major independent factors driving the randomness in the price.

Correlation exists between the prices of commodities. Hence, we model the randomness contributed by the principal components on the commodity prices are correlated.

As described, the general effect of each principal component on the price is random and thus in a simulation, a correlated random number is generated for each principal component. These random numbers are known as the Fourier coefficients. The final price for a given contract is then simply its seasonal mean plus the random Fourier coefficients multiplied by its corresponding principal components.

Each of the principal components are column vectors containing as many elements as contracts in the term structure. The magnitude of the elements of the vector corresponding to the earlier contracts, such as one, two, three, etc. month contracts are usually higher than later contracts, such as the 30-month contract, which increases the volatility of earlier contracts. This correctly models the observation that shorter contracts normally exhibit higher volatility than longer contracts.

Since the complex stochastic behavior in commodity prices cannot be fully described by one-factor models. Schwartz and Smith (2000) proposed a two-factor model that assumes the first factor to be a zero-mean Ornstein–Uhlenbeck process, representing short term fluctuations of price. The second factor is modeled by the arithmetic Brownian Motion, representing a long-term effect. Both factors are assumed to be correlated.

Borovkova and Geman (2006, 2007) introduces a synthetic factor model that prices options in a robust way. The approach approximates the distribution of a basket option based on the moment matching technique.

Ladokhin & Borovkova (2021) extends Borovkova and Geman (2007) into a three-factor model that incorporates the synthetic spot price. The model is based on liquidly traded futures and stochastic level of mean reversion.

In this article, we present a generic multi-factor model. The model clearly illustrates the essence of the multi-factor driving of commodity prices. The model describes the commodity price as a sum of mean reverting factors. Factors with low mean reversion rate are called Slow/Long Factors, and represent uncertainties on long-term fundamentals. Factors with high mean reversion rate are called Fast/Short Factors, and are used to model short term risk factors such as weather, middle-east unrest, economic slowdown, central-bank policy changes, etc.

We also propose a generic calibration procedure for the multi-factor model. The calibration procedure consists of an offline step where the mean reversion rates, the ratio of the long and short factor volatilities and the correlation between the long and short factors are determined via historical analysis. This offline step is performed relatively infrequently. There's also an online step of the calibration which happens every time the model is used to price an option or to simulate price paths.

The online calibration ensures that the model prices for vanilla options match exactly the market prices. Finally, there is an introspective step where trading desk adjusts the factor parameters in the offline step to better match the market prices for non-vanilla options such as swaptions and term-settled options. This introspective step happens whenever we have an observation of the market price for these illiquid options.

The rest of this article is organized as follows: First we describe the existing commodity factor models. Second, we present a new multi-factor model for commodity prices. Third, we propose a generic calibration method for the model. Then, we elaborate the valuation of commodity derivatives and Monto-Carlo simulation. Next, we discuss the empirical result. Finally, the conclusions are provided.

## 2. Single-, Two-, and Three-Factor Models

A single factor model assumes that the forward curve of an asset is driven entirely by the spot price. Thus, a single-factor model suffices for these types of assets. The generic formulation reduces to

$$\frac{dF^T(t)}{F^T(t)} = p(t)dW(t)$$

The function p(t) is the familiar time-dependent local volatility function in the equity derivatives modeling literature, which is typically calibrated via bootstrapping to the term structure of market implied volatility for options at different maturities.

Data analysis on historical crude oil forward and futures prices revealed that more than 90% of the variance of the forward curve can be captured by the first two principal components. Hence, it's customary to use a two-factor model to describe the dynamics of crude oil futures curve. Since crude oil is easy to store and globally traded, it is less subject to seasonal fluctuations than natural gas or electricity.

Crude oil futures with different delivery months are intrinsically linked. Therefore, people simplify the model by dropping the T-dependent volatility function q. To reduce the number of parameters, we also assume that the mean reversion rate of the long factor is zero, leaving only a single mean reversion rate for the short factor to be determined.

The model can be simplified as

$$\frac{dF^T(t)}{F^T(t)} = p_S(t)e^{-\beta(T-t)}dW_S(t) + p_L(t)dW_L(t)$$

In the limit of $T \gg t$, the short factor term drops out, and we see that the long factor is the sole driver for a long-dated futures contract. In the limit of $T = t$, the exponential term goes to 1, and we see that the short factor provides an extra "kick" to the prompt futures contract in addition to the long factor

To mention in passing, the same non-seasonal two-factor model is also used for refined oil products and base metals.

Natural gas and power are strongly seasonal, and such assets delivered at different months (or even at different hours of the day in the case of electricity) have distinct characteristics and can be thought of as different commodities. To reflect this in the model, we utilize the T-dependent volatility q and drop the t-dependent volatility p. The model reduces to:

$$\frac{dF^T(t)}{F^T(t)} = q_S(T)e^{-\beta(T-t)}dW_S(t) + q_L(T)dW_L(t)$$

For financial derivatives such as futures, it mostly suffices to use a two-factor model. For physical natural gas or power (or financial contracts that closely track physical price such as gas daily index), sometimes we need to add a "super"-fast factor to model the weather induced shocks. For example, in an extremely hot day, the air-conditioning usage may lead to a demand peak in power and the power price can suddenly jump by many folds. The usual two-factor model will not be able to capture this risk occurring only at the time scale of the delivery day or hour.

A three-factor model may also be useful to value the time spread options, particularly a portfolio of time spread options, such as natural gas storage contract. In the two-factor model, the two principal modes of motion are the parallel curve shift and the curve rotation. Time spread options are not sensitive to the primary mode of curve shift. They are sensitive to curve rotation and other curve shape changes beyond the first two principal components.

## 3. Multi-factor Model

Commodity derivatives typically trade by futures or forwards and thus there's a need to model the dynamics of the entire forward curve. The application of such a model includes semi-analytical pricing of non-vanilla options, Monte-Carlo simulation, risk management, etc. The Multi-factor model presented in this guide is capable of modeling a wide range of commodity assets, as well as FX and equities.

The model assumes that the forward curve is driven by N random factors, characterized by different time scales. Mathematically, the over-arching formula for the futures price $F^T$ reads

$$\frac{dF^T(t)}{F^T(t)} = \sum_{i=1}^{N} p_i(t) q_i(T) e^{-\beta_i(T-t)} dW_i\,(t)$$

Here, t is the current time, T is the futures expiration time, i is the index of the i'th factor, N is the total number of factors, $p_i$ is the local time dependent volatility of the i'th factor, $q_i$ is the expiry time dependent volatility of the i'th factor, $\beta_i$ is the mean reversion rate of the i'th factor, and $W_i$is a standard Brownian motion that is the random driver for the i'th factor. The Brown motions are correlated with the correlation matrix with element $\rho_{i,j}$, i.e. $\langle dW_i, dW_j \rangle = \rho_{i,j} dt$.

Here are several particular examples of the generic formulation above, from which the intuition of the model should come alive. But for now, let's expand certain properties of the generic model a bit more. First, we define the i'th Mean Reverting Factor $Y_i$ by the following OU process

$$dY_i(t) = -\beta_i Y_i(t) dt + p_i(t) dW_i(t)$$

We assume that the process starts from 0, i.e. $Y_i(0) = 0$. The solution to this SDE is easily derivable:

$$d[e^{\beta_i t} Y_i(t)] = e^{\beta_i t} p_i(t) dW_i(t)$$

$$Y_i(t) = \int_0^t e^{-\beta_i(t-s)}\, p_i(s) dW_i(s)$$

The futures price can be expressed in terms of the mean reverting factor as follows:

$$\frac{dF^T(t)}{F^T(t)} = \sum_{i=1}^{N} q_i(T)e^{-\beta_i T} d[e^{\beta_i t} Y_i\,(t)]$$

$$F^T(t) = F^T(0)e^{\sum_{i=1}^{N} q_i(T)e^{-\beta_i(T-t)} Y_i(t) - \frac{1}{2}\langle \log F^T \rangle_0^t}$$

Here, $\langle \log F^T \rangle_0^t$ is the quadratic variation of the log futures price from time 0 to time t. The significance of the formula above is that the futures price can be explicitly put in a simple algebraic formula involving the N mean reverting factors which do not depend on T. This means that, in a Monte-Carlo setting, the task of simulating the entire forward curve boils down to that of simulating just the N mean reverting factors.

The quadratic variation is a specific case of a very useful quantity: Log Covariance. We define the log covariance of two futures contracts, $F^{T_1}$ and $F^{T_2}$, integrating from $t_1$ to $t_2$ as

$$\langle \log F^{T_1}, \log F^{T_2} \rangle_{t_1}^{t_2} = \sum_{i,j=1}^{N} q_i(T_1)q_j(T_2) \int_{t_1}^{t_2} p_i(s)p_j(s)\, e^{-\beta_i(T_1-s)-\beta_j(T_2-s)} \rho_{i,j} ds$$

In particular,

$$\langle \log F^T \rangle_0^t = \sum_{i,j=1}^{N} q_i(T)q_j(T) \int_0^t p_i(s)p_j(s)\, e^{-(\beta_i+\beta_j)(T-s)} \rho_{i,j} ds$$

The value of a vanilla option is computed by substituting the volatility $\sqrt{\frac{\langle \log F^T \rangle_0^t}{t}}$ into the Black-Scholes formula.

Substituting T=t into the equation above gives the formula for the spot price of a commodity:

$$S(t) = S(0)e^{\sum_{i=1}^{N} q_i(t)Y_i(t)+\cdots}$$

This equation most clearly illustrates the essence of the multi-factor model. The model describes the (log) commodity price as a sum of mean reverting factors. Factors with low mean reversion rate are called Slow/Long Factors, and represent uncertainties on long-term fundamentals. Factors with high mean reversion rate are called Fast/Short Factors, and are used to model short term risk factors such as weather, middle-east unrest, economic slowdown, central-bank policy changes, etc.

Finally, if we have two assets that are modeled by the multi-factor model, with potentially different number of factors, we can calculate the Cross-Asset Log Covariance as follows.

$$\frac{dF_1^T(t)}{F_1^T(t)} = \sum_{i=1}^{N_1} p_{i,1}(t) q_{i,1}(T) e^{-\beta_{i,1}(T-t)} dW_{i,1}(t)$$

$$\frac{dF_2^T(t)}{F_2^T(t)} = \sum_{j=1}^{N_2} p_{j,2}(t) q_{j,2}(T) e^{-\beta_{j,2}(T-t)} dW_{j,2}(t)$$

$$\langle \log F_1^{T_1}, \log F_2^{T_2} \rangle_{t_1}^{t_2} = \sum_{i=1}^{N_1} \sum_{j=1}^{N_2} q_{i,1}(T_1) q_{j,2}(T_2) \int_{t_1}^{t_2} p_{i,1}(s) p_{j,2}(s)\, e^{-\beta_{i,1}(T_1-s)-\beta_{j,2}(T_2-s)} \rho_{i,j}^{1,2} ds$$

Here, $\rho_{i,j}^{1,2}$ is the instantaneous correlation between the i'th factor of the first asset and the j'th factor of the second asset, i.e.

$$\rho_{i,j}^{1,2} dt = \langle dW_{i,1}(t), dW_{j,2}(t) \rangle$$

The cross-asset log covariance is useful in pricing basket option or foreign currency denominated options.

The standard multi-factor model specifies a log-normal process for the futures price, and therefore does not incorporate volatility smile directly. The model needs to be adjusted to handle the volatility smile. There are several sophisticated approaches to do this. However, for most non-vanilla options, the following simpler methodology gives satisfactory results:

i. Price the non-vanilla option using the multi-factor model calibrated to the term structure of ATM volatilities for vanilla options. We call the volatility of the non-vanilla option the composite volatility.

ii. Compute the "moneyness" of the non-vanilla option using the composite ATM volatility obtained in step 1.

iii. Now construct a new multi-factor model, this time calibrated to the term structure of volatilities for vanilla options at that moneyness.

iv. Reprice the non-vanilla option using the new Multi-factor model, which gives the final price of this option.

The most convenient "moneyness" metric is the so-called **Quick Delta** (QD). QD measures how many standard deviations away the forward price is relative to the strike of the option, in the log-normal space. QD is an approximation to the Black-Scholes (put) delta. But it is more convenient than Black-Scholes delta for the purpose of parameterizing the volatility surface because QD for ATM option is exactly 50% and it is easy to convert QD to a dollar strike or the other way around. Mathematically, QD is defined as

$$\text{QD} = N(\log\frac{\text{Strike}}{\text{Forward}}/\sigma\sqrt{T})$$

Here, N is the cumulative normal density function, $\sigma$ is the ATM volatility, and T is the option expiration. The cumulative normal density function is applied to transform the moneyness to a number between 0 and 1.

As an example, let's apply the procedure above to swaption pricing.

i. Price the swaption using the Multi-factor model calibrated to the term structure of ATM volatilities for vanilla options. We obtain the ATM log variance $\mathbb{V}_{ATM}[\log F]$ of the swap rate $F = \sum_{k=1}^{M} w_k F^{T_k}(t_k)$ . The ATM swaption volatility is then $\sigma_{ATM\ Swaption} = \text{Sqrt}(\mathbb{V}_{ATM}[\log F]/T_{Swaption})$.

ii. Compute the "moneyness", i.e. quick delta of the swaption

$$\text{QD} = N(\log\frac{\text{Swaption Strike}}{\text{Swap Rate F}}/\sigma_{ATM\ Swaption}\sqrt{T_{Swaption}})$$

iii. Now construct a new multi-factor model, this time calibrated to the term structure of volatilities for vanilla options at that QD. For example, if QD = 25%, we will take the volatility at 25 QD for all the vanilla options, and recalibrate a new multi-factor model using that set of volatilities.

iv. Reprice the swaption using the new multi-factor model. This gives the final price of the swaption.

## 4. Model Calibration

The goals of the model calibration are to achieve: i) the market prices of vanilla options need to be perfectly matched by the model; ii) the model prices for non-vanilla options are in good agreement with the market prices; and iii) the implied model dynamics are in good agreement with the characteristics of the historical data series. We obtain data from FinPricing (2020).

Since the historical analysis is backward looking and option pricing should be forward looking and marked to market, goal 2 takes precedence over goal iii)

In this section, we show how the two-factor model parameters are determined in the offline step of the calibration. The purpose of the offline step is to back out the mean reversion rate ($\beta$), the volatility ratio ($p_S/p_L$for non-seasonal model and $q_S/q_L$ for seasonal model), and the correlation ($\rho$) between the long and short factors from the historical time series. There are of course many ways of studying the historical data and it is clearly beneficial to benchmark the model dynamics against history in multiple ways. I'll show 3 different methods here.

The first method is Principal Component Analysis. PCA is a mathematical procedure that uses an orthogonal transformation to convert a set of observations of possibly correlated variables into a set of values of linearly uncorrelated variables called Principal Components. The number of principal components is less than or equal to the number of original variables.

This transformation is defined in such a way that the first principal component has the largest possible variance (that is, accounts for as much of the variability in the data as possible), and each succeeding component in turn has the highest variance possible under the constraint that it be orthogonal to (i.e., uncorrelated with) the preceding components. Principal components are guaranteed to be independent if the data set is jointly normally distributed.

We apply PCA to the historical daily returns of the futures curve and compare the first two principal components with those implied by the two-factor model. The mean reversion rate, volatility ratio and correlation are chosen to achieve the best fit to the historical PCA in the least squared sense.

The following graphs illustrate the quality of fitting for West Texas Intermediate (WTI). The historical period is between July 24, 2011 and July 24, 2013, with more recent data given higher importance. The weighting of the data follows an exponential formula with the decay rate corresponding to a half life of 125 business days. The two-factor model parameters are 0.35 for mean reversion rate, 1.6 for volatility ratio and -20% for correlation.

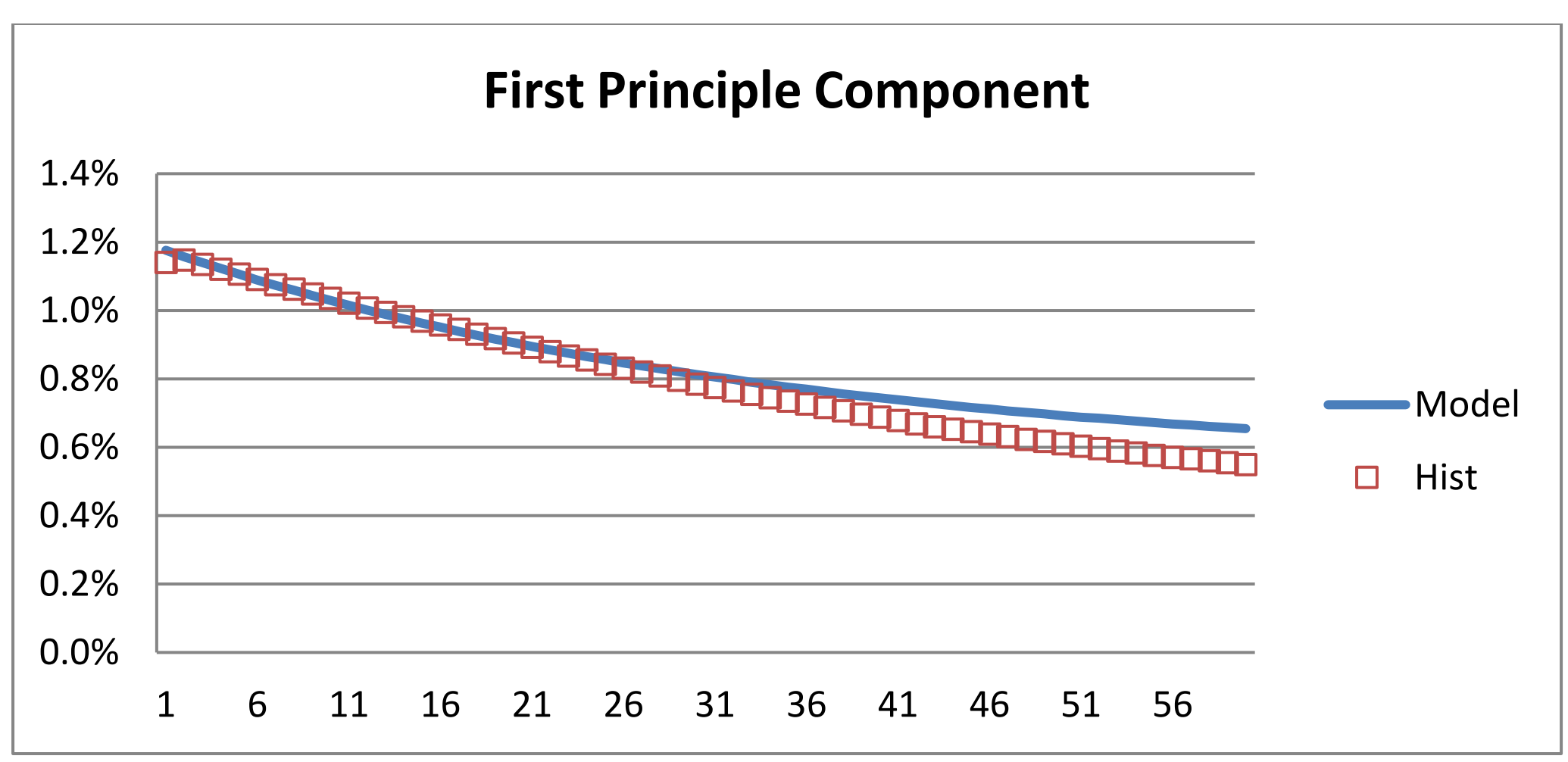

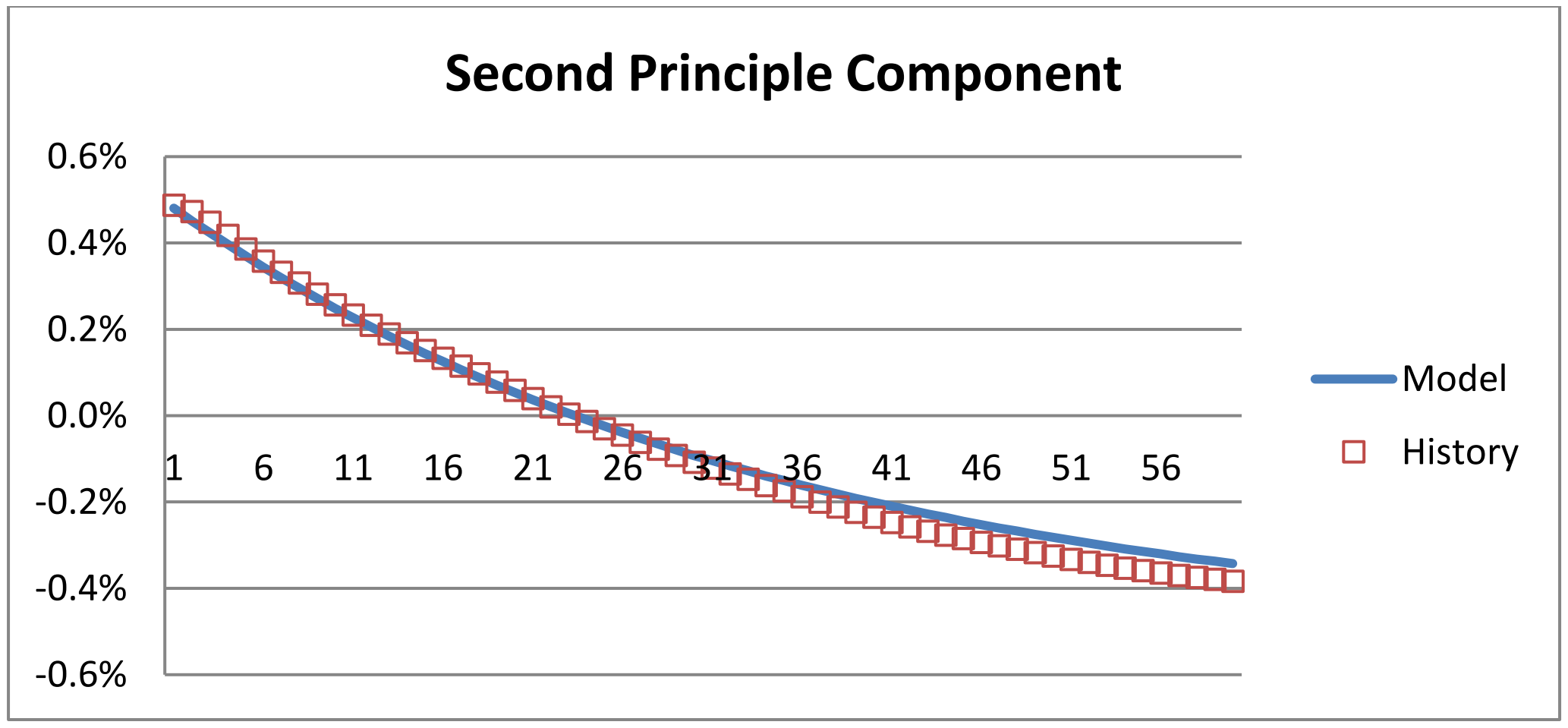

FIGURE 1 illustrates the quality of fitting for West Texas Intermediate (WTI).

The second method is to directly compare certain statistical metrics between the historical data and the model. As far as the most liquid non-vanilla options are concerned, the most important statistics are the relative volatilities of different futures and the correlations between these futures. The figures below illustrate the quality of fitting in terms of these two metrics. The historical data and two-factor model parameters are the same as above.

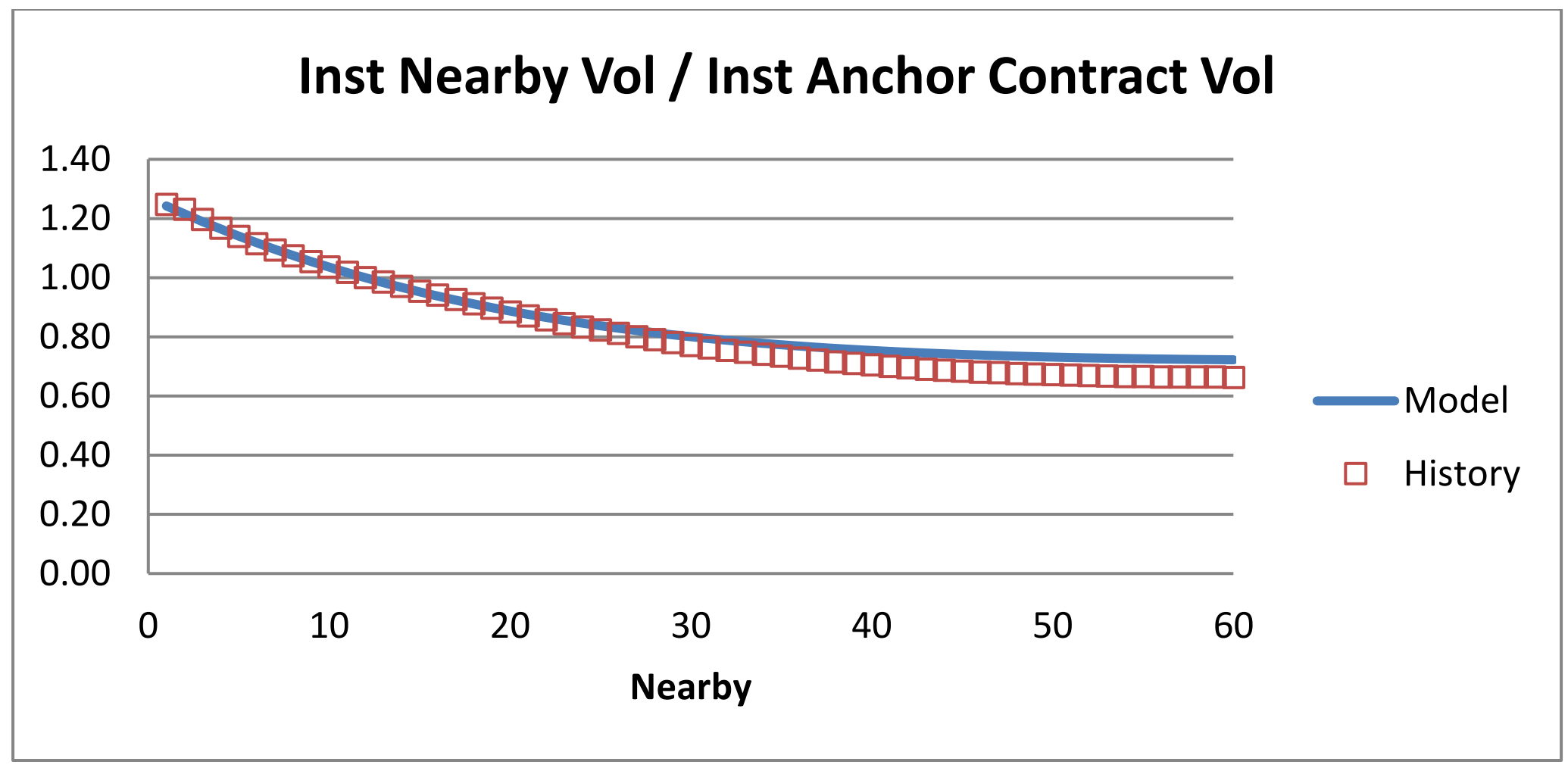

FIGURE 2 illustrates the relative volatilities of different futures for West Texas Intermediate (WTI). More specifically, the graph plots the ratio of the instantaneous volatility of the nearby number in the x-axis versus the instantaneous volatility of the 12th nearby.

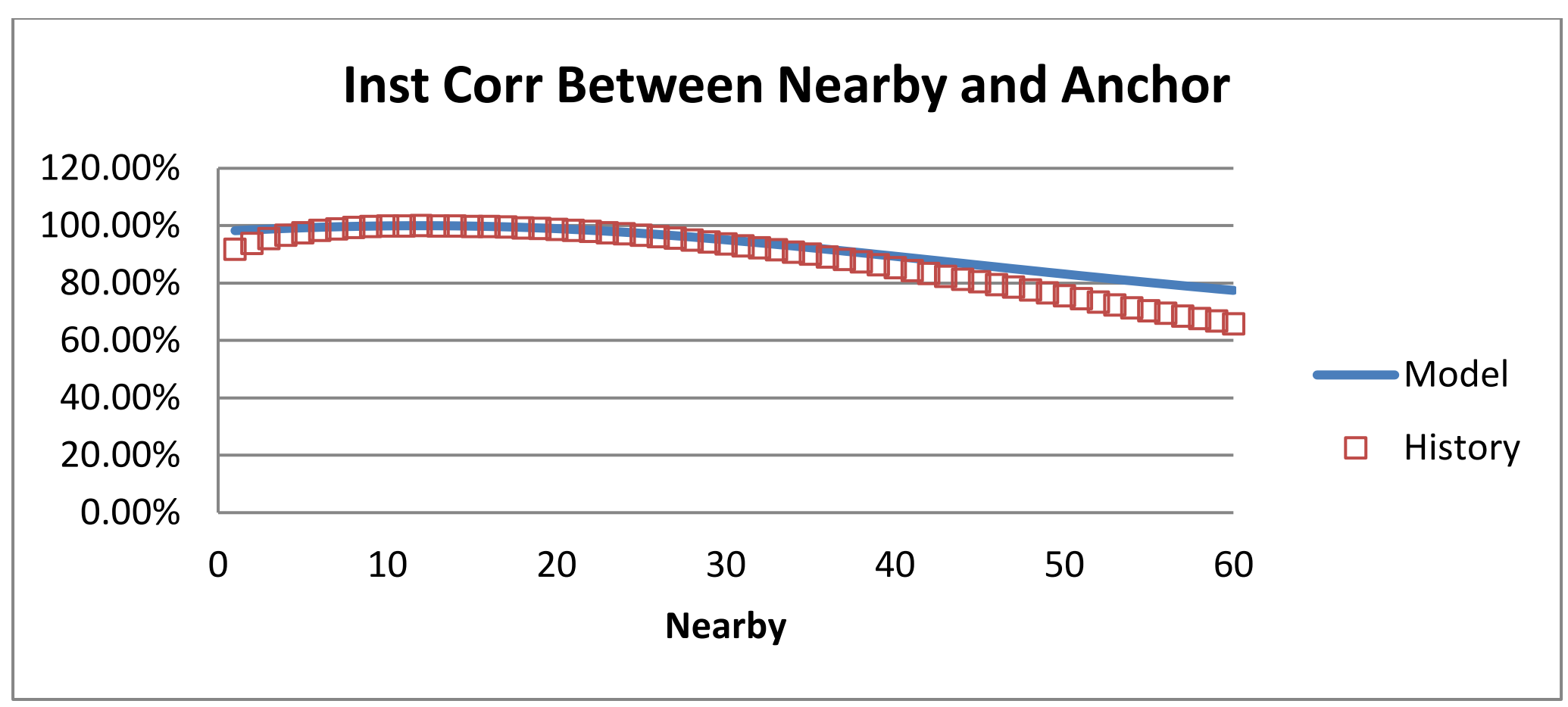

FIGURE 3 illustrates the correlation between different futures for West Texas Intermediate (WTI). More specifically, the graph plots the instantaneous correlation of the nearby number in the x-axis and the $12^{th}$ nearby.

The third method is so-called relative value analysis. Both the first and second methods are purely historical analysis. The strength of the relative value analysis is that it combines information from historical analysis and information from market implied volatilities of the vanilla options.

The key quantity that links the history and the future is the ratio between the realized swap volatility and the realized futures volatility. For example, if in calendar 2013, the realized volatility ratio between Cal14 swap and the Feb14 future contract was 0.9. Suppose on Jan 2, 2014, we need to price a Cal15 swaption and the Feb15 vanilla option is priced at 20% implied volatility.

The relative value analysis would suggest that we price the Cal15 swaption using 20% * 0.9 = 18% volatility. This gives us a price target for the swaption. We can establish the price targets for a number of tenors, Cal16, Cal17, 2H14, 1H15, etc., and then adjust the mean reversion, volatility ratio and correlations to match as well as possible these target prices. Here's an example of doing historical relative value analysis on Cal14 swap using the historical period of 2013.

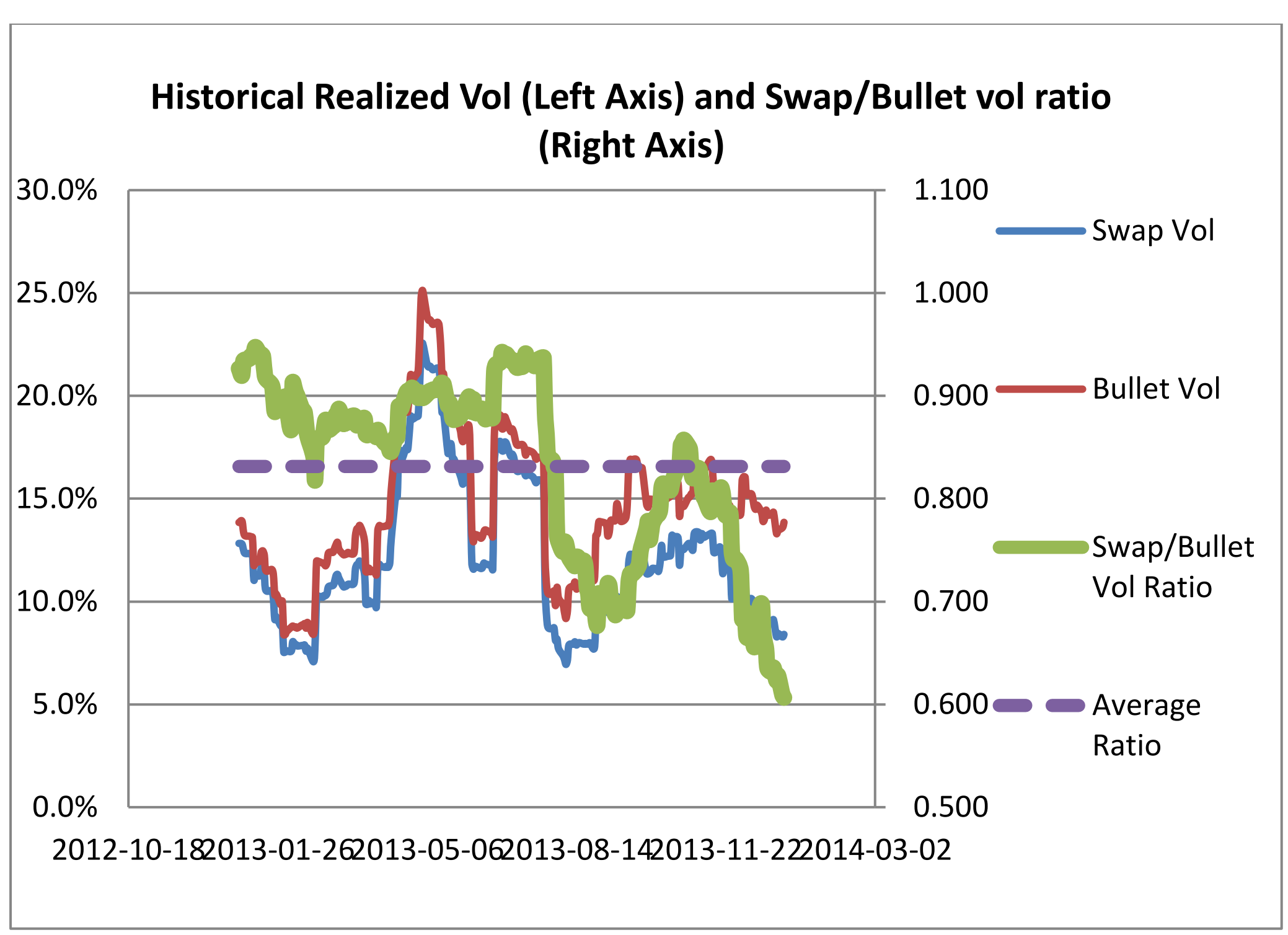

FIGURE 4 shows historical realized vol vs. swap vol ratio

Note that the two-factor model parameters we backed out from historical analysis are not used directly to price new deals or manage our risks. Historical analysis is backward looking, while derivatives pricing should reflect our forward looking view. Therefore, we use the introspective step to complement the historical analysis with the information embedded in the market non-vanilla option prices about how the future will hold.

We use the historical two-factor model parameters to price non-vanilla options such as swaptions or term-settled Asian options, and fine tune the parameters so that the model price is in good agreement with the market prices (these non-vanilla trades typically have wide bid-offer in the market so we don't need to match the market price exactly).

Once the mean reversion rates, the volatility ratio and the correlations are fixed, the only thing left is to impute the overall level of the factor volatilities. The purpose of the online step of the calibration is to solve for the factor volatilities such that the model prices vanilla options at exactly their market prices. We now return to the general Multi-factor model setting to lay out the procedure of the online calibration. With the fixed volatility ratio, we can express the t-dependent factor volatilities in terms of the constant volatilities vector and a t-dependent scaling factor $\alpha(t)$, i.e.

$$p_i(t) = p_i\alpha(t)$$

We assume the scalar function $\alpha(t)$ is a piecewise constant function, with the time knots being the standard option expiration dates $t_1, t_2, t_3, \ldots$. Namely,

$$\alpha(t) = \alpha_k, \quad \text{for} \;\; t_{k-1} \le t < t_k \text{ (where } t_0 = 0)$$

The log variance of a futures contract with maturity $T_M$ up to its standard option expiry $t_M$ reduces to

$$\langle \log F^{T_M} \rangle_0^{t_M} = \sum_{i,j=1}^{N} \int_0^{t_M} p_i(s)p_j(s)\, e^{-(\beta_i+\beta_j)(T_M-s)}\rho_{i,j}ds$$

$$= \sum_{i,j=1}^{N} p_i p_j \rho_{i,j} e^{-(\beta_i+\beta_j)T_M} \sum_{k=1}^{M} \alpha_k^2 \frac{e^{(\beta_i+\beta_j)t_k} - e^{(\beta_i+\beta_j)t_{k-1}}}{\beta_i+\beta_j}$$

We now use a bootstrapping procedure to solve for $\alpha_k$ for k=1, 2 … .The price of the first option (with the nearest expiration) depends only on $\alpha_1$, and so $\alpha_1$ can be backed out from the option price. The price of the second option depends on both $\alpha_1$ and $\alpha_2$. Since $\alpha_1$ is already determined, we just need to solve for $\alpha_2$ to match the second option. We continue this procedure to solve for the values for all $\alpha_k$.

The instruments used in this calibration are vanilla options on futures, i.e., vanilla options expiring in $t_M$ on the futures expiring on $T_M$. The vanilla option volatilities used in the calibration are marked by the trading desk.

The difference with the seasonal model is to use T-dependent volatility functions. Analogous to the non-seasonal model, the factor volatility functions can be expressed as a constant vector multiplied by a T-dependent scalar, namely:

$$q_i(T) = q_i\lambda(T)$$

The log variance of a futures contract with maturity $T_M$ up to the its standard option expiry $t_M$ is equal to

$$\langle \log F^{T_M} \rangle_0^{t_M} = \sum_{i,j=1}^{N} \int_0^{t_M} q_i(T_M) q_j(T_M)\, e^{-(\beta_i+\beta_j)(T_M-s)} \rho_{i,j} ds$$

$$= \lambda^2(T_M) \sum_{i,j=1}^{N} q_i q_j \rho_{i,j} e^{-(\beta_i+\beta_j)T_M} \frac{e^{(\beta_i+\beta_j)t_M} - 1}{\beta_i+\beta_j}$$

This is easier to solve than the non-seasonal model and no bootstrapping procedure is required.
The instruments used in this calibration are vanilla options on futures, i.e., vanilla options expiring in $t_M$on the futures expiring on $T_M$. The vanilla option volatilities used in the calibration are marked by the trading desk.

As mentioned earlier, the model parameters backed out from the historical analysis are not directly used to value our book and manage the risks. Option pricing needs to be forward looking. The historical analysis is useful in that it gives us a sense of "value" (are we stepping into an overheated market?), but no amount of historical analysis is going to tell us for sure what is going to happen in the future. So we will have to adjust the model parameters such that the swaption prices are close to the market prices. This section describes how this last step of the calibration works.

There are two main sources of swaption market prices that we use to gauge how other deals are valuing the swaptions. The first source is the inter-dealer broker market. The second source is the monthly Totem swaption valuation service that provides a consensus (average) price among the participants of this service. Both sources are valuable, and neither is perfect. The broker quotes are typically spotty, i.e., we may get a quote once in a few days or sometimes weeks. And the broker quotes only reflect the view of one dealer who is quoting. The third problem with deal quotes is that dealers tend to skew the bids lower since the market flow is essentially one way (dealers are buying swaption from clients).

The value of broker quotes however is that they are transact-able. On the other hand, Totem consensus gives us a way to regularly check our valuation against a group of dealers. But it is not perfect because some of the participants may not be active dealers in the swaption market, and their prices can be less reliable than other dealers. Another problem with Totem is that the prices are only what dealer use to mark their book, but not necessarily transact-able.

In commodities, our current profit and loss (PnL) policy on swaptions is to use Totem as the official MTM for each month end. This is implemented by comparing the model price and Totem consensus at each month end and adjusting the official PnL by the difference. As a result, we would like to mark the model to be "close" to Totem prices. We use quotation mark on "close" because we do not want to match the Totem prices exactly, given the limitations of the Totem data noted earlier. It is better for us to also use broker quotes as a supplement. If broker quotes are noticeably lower than Totem, we should in general mark the model conservative to Totem. If the broker quotes are noticeably higher than Totem, it is OK to mark the model slightly aggressive to Totem.

In practice, we would like to hold the mean reversion, volatility ratio and correlation parameters as steady as possible. Changing these parameters will change the Greeks, and requires re-balancing of the futures and options hedges, hence incurring costs. Therefore, we will only change these parameters when they noticeably deviate from both Totem and broker quotes. More specifically, we consider re-calibration if the model price is more aggressive than both Totem and broker offers. We also consider re-calibration if the model price is more conservative than Totem and broker bids.

Let me use a real-life example to illustrate this process. At Feb 2014 month end, we have the following comparison between the model price and Totem consensus for WTI swaptions. The model prices are within 1% of Totem for Cal17, 18, and 19. For cal15 and 16, the model price is conservative by more than 1%. Shall we recalibrate the model to show higher prices for Cal15 and 16?

| Straddle Swaption | Model Price | Totem Consensus | Diff % |
|---|---:|---:|---:|
| Cal15 | 9.49 | 9.97 | -4.8% |
| Cal16 | 12.86 | 13.07 | -1.6% |
| Cal17 | 15.07 | 15.13 | -0.4% |
| Cal18 | 16.69 | 16.71 | -0.1% |
| Cal19 | 17.87 | 17.8 | 0.4% |

Table 1: The comparison of the model price and market price for WTI swaptions

In the next few weeks, we got the following broker quotes and compared with our model. Clearly, our model is higher than the broker bids. So we are fine with the model being conservative relative to Totem, knowing that we are not over-conservative (which would be the case if our model is even lower than the broker bid). No re-calibration is needed.

| Instrument | Quote Date | Ref Swap | Bid | Offer | Model Price |
|---|---|---|---|---|---|
| Cal15 ATM Call | 3/19/2014 | 88.00 | 4.57 | 4.97 | 4.75 |
| Cal16 ATM Call | 3/19/2014 | 83.50 | 6.05 | 6.45 | 6.49 |
| Cal17 ATM Call | 3/24/2014 | 81.75 | 6.93 | 7.33 | 7.42 |
| Cal18 ATM Call | 3/24/2014 | 80.50 | 7.60 | 8.10 | 8.22 |
| Cal19 ATM Call | 3/24/2014 | 80.00 | 8.26 | 8.76 | 8.87 |

Table 2: The comparison between broker quotes and model prices.

Finally, if the model needs to be recalibrated, we tune the model parameters (mean reversion, vol ratio, and correlation) to put the swaption prices back in the acceptable (as determined by the procedure above) range again. Please refer to the following table for the sensitivity to the model parameters. Tweaking the mean reversion higher has an impact of reducing the swaption price in the front tenors, and increasing the swaption price in the back tenors.

Tweaking the vol ratio higher has an impact of lowering the swaption price across the board. Tweaking the correlation higher reduces the swaption value across all tenors, but the impact is higher for the swaptions in the back. One can use the 3 parameters to achieve different "shape" changes. For example, if one wants to lower the swaption values by similar amount across tenor. The most effective tool is to increase the vol ratio. Mean reversion and correlation provide further flexibility for tuning the "shape".

| ATM Call | Base Price | Mean Rev + 0.1 | Diff to Base | Vol Ratio + 0.1 | Diff to Base | Corr + 10% | Diff to Base |
|---|---|---|---|---|---|---|---|
| Cal15 | 4.746 | 4.724 | (0.022) | 4.688 | (0.059) | 4.730 | (0.016) |
| Cal16 | | | | | | | |

| | 6.430 | 6.428 | (0.002) | 6.358 | (0.072) | 6.389 | (0.041) |
|---|---|---|---|---|---|---|---|
| Cal17 | 7.533 | 7.542 | 0.009 | 7.460 | (0.073) | 7.478 | (0.055) |
| Cal18 | 8.347 | 8.360 | 0.013 | 8.277 | (0.070) | 8.287 | (0.060) |
| Cal19 | 8.934 | 8.949 | 0.015 | 8.870 | (0.065) | 8.873 | (0.061) |

Table 3: The impact of parameters on prices

## 5. Pricing Commodity Derivatives via Multi-Factor Model

Asian option settles on the average of the futures price over a sampling period. Typically, the prompt future is used as the sampling contract. Since the arithmetic average of the lognormal variables is no longer lognormal, one typically utilizes some approximation to price Asian options. The most common method is to assume that the arithmetic average is still a lognormal variable and to use moment matching to determine the variance of this average.

Suppose the Asian option has M sampling dates $t_1, t_2, \dots, t_M$, with sampling weights $w_1, w_2, \dots, w_M$. The average price for which we need to calculate the log variance is the following:

$$F = \sum_{k=1}^{M} w_k F^{T_k}(t_k)$$

By the moment matching approximation, the log variance of F can be solved as

$$\mathbb{V}[\log F] = \log \frac{\mathbb{E}[F^2]}{\mathbb{E}^2[F]}$$

The expected value in the numerator can be further expanded as follows

$$\mathbb{E}[F^2] = \sum_{j,k=1}^{M} w_j w_k \mathbb{E}\left[F^{T_j}(t_j) F^{T_k}(t_k)\right]$$

$$= \sum_{j,k=1}^{M} w_j w_k \mathbb{E}\left[F^{T_j}(t_j)\right] \mathbb{E}\left[F^{T_k}(t_k)\right] e^{\langle \log F^{T_j}, \log F^{T_k} \rangle_0^{\mathrm{Min}(t_j,t_k)}}$$

Finally, the log variance is plugged into the Black-Scholes equation to arrive at the Asian option price.

Another example is commodity swaption. A swaption is similar to Asian option as it is also an option on the average of futures prices. The difference is that a swaption expires before the averaging period starts. The formula to price a swaption using the Multi-factor model is very similar to the Asian option formula above, except that the sampling dates $t_1, t_2, \ldots, t_M$ are all the same and equal to the swaption expiration date.

Moreover, for OTC swaptions, the discount factors from the swap settlement dates to the swaption expiration date are baked into the sampling weights. More specifically, if the undiscounted sampling weights for the underlying futures are $(u_1, u_2, \ldots, u_M)$, and the discount factors from the swaption expiration date to the settlement dates of each underlying swaplet are $(d_1, d_2, \ldots, d_M)$, then the discounted sampling weights which are used in swaption pricing are

$$(w_1, w_2, \ldots, w_M) = (u_1 * d_1, u_2 * d_2, \ldots, u_M * d_M)$$

So far, we have covered the pricing formula for a native denominated (USD typically) Asian option or Swaption. We now discuss how to price Asian or Swaption denominated in a foreign currency. To this end, we first review the stochastic calculus associated with a foreign currency.

Let's denote by d the domestic economy (typically USD), and by f the foreign economy. Let's use $P$ to stand for the domestic risk-free measure, and $\tilde{P}$ for the foreign measure. X(t) is the spot FX rate at time t to converts 1 unit of foreign currency into the domestic currency. Y(t) = 1/X(t) is the spot FX rate that converts 1 unit of the domestic currency to the foreign currency. As usual, $F^T(t)$ denotes the futures price in the domestic currency, and we define $G^T(t) = F^T(t)Y(t)$ to be the futures price in the foreign currency.

The dynamics of X(t) is governed by the following SDE, where the Brownian motion W is in the domestic measure $P$,

$$\frac{dX(t)}{X(t)} = \left(r_d(t) - r_f(t)\right) dt + \sigma_X(t) dW(t)$$

Here, $r_d(t)$ and $r_f(t)$ are the deterministic interest rates for the domestic and foreign currencies respectively. By Ito lemma,

$$dY(t) = -\frac{dX}{X^2} + \frac{\langle dX \rangle}{X^3} = Y(t) \left[ \left(r_f(t) - r_d(t)\right) dt - \sigma_X(t)(dW(t) - \sigma_X(t)dt) \right]$$

By the Girsanov theorem, we can rewrite this SDE in the foreign measure $\tilde{P}$, where $d\widetilde{W}(t) = dW(t) - \sigma_X(t)dt$ is a Brownian motion in the foreign measure,

$$\frac{dY(t)}{Y(t)} = \left(r_f(t) - r_d(t)\right) dt - \sigma_X(t) d\widetilde{W}(t)$$

Therefore, Y and X have the same volatility and covariance.

Consider an Asian option that settles on the average price in the foreign currency:

$$G = \sum_{k=1}^{M} w_k G^{T_k}(t_k) = \sum_{k=1}^{M} w_k F^{T_k}(t_k) Y(\, t_k)$$

It is convenient to price the foreign option in the foreign measure $\tilde{P}$, and we use the same moment matching method as the native option,

$$\mathbb{V}[\log G] = \log \frac{\mathbb{E}_{\tilde{P}}[G^2]}{\mathbb{E}_{\tilde{P}}{}^2[G]}$$

The first moment is easy to compute

$$\mathbb{E}_{\tilde{P}}[G] = \sum_{k=1}^{M} w_k \mathbb{E}_P[F^{T_k}(t_k)] \mathbb{E}_{\tilde{P}}[Y(t_k)]$$

The first expectation in the right-hand side of the equation is simply today's futures price, and the second expectation is today's forward FX rate.

The second moment can be expanded as

$$\mathbb{E}_{\tilde{P}}[G^2] = \sum_{j,k=1}^{M} w_j w_k \mathbb{E}_{\tilde{P}}\left[G^{T_j}(t_j) G^{T_k}(t_k)\right]$$

$$= \sum_{j,k=1}^{M} w_j w_k \mathbb{E}_{\tilde{P}}\left[G^{T_j}(t_j)\right] \mathbb{E}_{\tilde{P}}\left[G^{T_k}(t_k)\right] e^{\langle \log G^{T_j}, \log G^{T_k} \rangle_0^{\mathrm{Min}(t_j,t_k)}}$$

The expected values are computed the same way as in the first moment. What remains is to calculate the log-covariance term:

$$\langle \log G^{T_j}, \log G^{T_k} \rangle_0^{\mathrm{Min}(t_j,t_k)}$$

$$= \langle \log F^{T_j}, \log F^{T_k} \rangle_0^{\mathrm{Min}(t_j,t_k)} + \langle \log F^{T_j}, \log Y \rangle_0^{\mathrm{Min}(t_j,t_k)} + \langle \log Y, \log F^{T_k} \rangle_0^{\mathrm{Min}(t_j,t_k)} + \langle \log Y, \log Y \rangle_0^{\mathrm{Min}(t_j,t_k)}$$

Since the FX process is modeled as a single-factor model, the log-covariance terms in the right-hand equation are a particular case of the cross-asset log-covariance formula.

The formula to price a foreign denominated swaption using the multi-factor model is very similar to the Asian option formula above, except that the sampling dates $t_1, t_2, \dots, t_M$ are all the same and equal to the swaption expiration date. Moreover, for OTC swaptions, the discount factors (for the foreign currency since we are pricing in the foreign measure) from the swap settlement dates to the swaption expiration date are baked into the sampling weights ($w_1, w_2, \dots, w_M$).

## 6. Multi-Factor Model Simulation

We have already seen in the multi-factor model, the task of simulating the entire forward curve boils down to that of simulating N mean reverting factors Y. In this section, we describe the details of the numerical algorithm to generate paths for these N mean reverting factors. With these mean reverting factors, the futures prices can be recovered easily using the algebraic formula above.

Suppose we want to simulate the mean reverting factors on M sampling dates $t_1, t_2, \ldots, t_M$, and we have already generated the paths up to time $t_{k-1}$. To simulate the next time step, we note that

$$Y_i(t_k) = e^{-\beta_i(t_k - t_{k-1})} Y_i(t_{k-1}) + \int_{t_{k-1}}^{t_k} e^{-\beta_i(t_k - s)} p_i(s) dW_i(s)$$

The N-by-N covariance matrix of the integral term is easily derivable from the Ito isometry:

$$\langle \int_{t_{k-1}}^{t_k} e^{-\beta_i(t_k - s)} p_i(s) dW_i(s), \int_{t_{k-1}}^{t_k} e^{-\beta_j(t_k - s)} p_j(s) dW_j(s) \rangle$$

$$= \int_{t_{k-1}}^{t_k} e^{-(\beta_i + \beta_j)(t_k - s)} p_i(s) p_j(s) \rho_{i,j} ds$$

We then calculate the Cholesky decomposition of the covariance matrix. Equipped with a random number generator capable of drawing N independent Gaussian random numbers, we can simulate the integral term and hence generate the paths for the mean reverting factors at the next time step $t_k$.

We may follow the recipe above in a multi-asset simulation (such as in CVA). It works theoretically, however in practice this requires calculating the full covariance matrix and its Cholesky decomposition across all assets and all factors at every time step. A more efficient algorithm goes as follows:

i. Assuming the correlation between the Brownian motions driving the i'th factor of the a asset and the j'th factor of the b asset is $\rho_{i,j}^{a,b}$, we compute the Cholesky decomposition of the full correlation matrix across all asset factors.

ii. With the Cholesky decomposition in step 1, one can easily generate the standard Gaussian random variables $R_i^a$ for all assets (indexed by a) and factors (indexed by i), such that $R_i^a$ and $R_j^b$ are correlated with the correlation $\rho_{i,j}^{a,b}$.

iii. For each asset a, take the random variables $R_i^a$ for all its factors, and reproduce a set of un-correlated random variables $Z_i^a$ by multiplying with the inverted Cholesky decomposition of the sub-correlation-matrix for the asset a. Notice that the sub-correlation-matrix is much smaller in dimensionality than the full correlation matrix (so this step is cheap).

iv. For each single asset a, we follow the recipe for single-asset simulation. Specifically, we compute the covariance matrix of the integral term at every time step, and use its Cholesky decomposition combined with the un-correlated random variables $Z_i^a$ to simulate the integral term at each time step.

v. Finally, step 4 allows us to advance the mean reverting factors of all assets in time. With the algebraic relation between the futures contracts and mean reverting factors, we obtain the futures price paths for all assets.

To make the multi-asset algorithm complete, we need to discuss how to determine the pairwise correlation $\rho_{i,j}^{a,b}$ between the i'th factor of the asset and the j'th factor of the b asset (where a could be equal to b, in which case the correlation is between different factors of the same asset). Obviously, one cannot hope to assign arbitrary values to these correlations and still get a semi-positive-definite (SPD) matrix that can be Cholesky decomposed. The simplest way to ensure SPD is to compute the correlation matrix directly from historical data.

But where do we get historical data on the factors? The factors are a model construct and are not directly observable from the historical data. What we observe is the historical futures prices, and we need to somehow "transform" them into factors. Here's a simple way to do this.

Take the two-factor model for WTI as an example. We pick two nearby contracts, one from the front end of the curve, say the 3rd nearby, i.e., $T_1 = t + 3/12$, and other from the back end of the curve, say the 36th nearby, i.e., $T_2 = t + 36/12$. According to the model,

$$\frac{dF^{T_1}(t)}{F^{T_1}(t)} = \sigma(t)[p_S e^{-\beta(T_1 - t)} dW_S(t) + p_L dW_L(t)]$$

$$\frac{dF^{T_2}(t)}{F^{T_2}(t)} = \sigma(t)[p_S e^{-\beta(T_2-t)} dW_S(t) + p_L dW_L(t)]$$

The left-hand sides of the equations, the log returns of the two futures, are directly observable from the historical time series. We can solve for the factors as follows:

$$p_S \sigma(t) dW_S(t) = \frac{\frac{dF^{T_1}(t)}{F^{T_1}(t)} - \frac{dF^{T_2}(t)}{F^{T_2}(t)}}{e^{-\beta(T_1-t)} - e^{-\beta(T_2-t)}}$$

$$p_L \sigma(t) dW_L(t) = \frac{e^{-\beta(T_2-T_1)} \frac{dF^{T_1}(t)}{F^{T_1}(t)} - \frac{dF^{T_2}(t)}{F^{T_2}(t)}}{e^{-\beta(T_2-T_1)} - 1}$$

Note that in this formula, the mean reversion rate is already determined by the procedure in Model Calibration. From this, the historical correlation between any two factors (either from the same asset or different assets) can be calculated.

## 7. Empirical and Numerical Study

Let's go through an example to understand the consequence of choosing the calibration strategy. At the time of this writing, we are in October 2013. Suppose we need to price an option on the December 2015 futures contract, while the option expires on the option expiration of the December 2014 contract. This is what we call an Early Expiry Option, which is a mini-swaption. T

The key element in pricing this option is the early expiry volatility of the December 2015 contract during the period from now to the expiration of the December 2014 contract. If we use a seasonal model to price this option, the early volatility of the option depends entirely on the trader's volatility mark for the December 2015 contract. If we use a non-seasonal model to price this option, the early volatility depends entirely on the trader's volatility mark for December 2014 contract. In other words, the Vega of the option would be attributed to different contract months depending on which calibration strategy to use.

Now let's imagine a market scenario where the implied volatility of the December 2014 contract increased by 2%, while the volatility of the December 2015 stays put. If we use the seasonal model, the Vega PnL under this market scenario would be flat. But if we use the non-seasonal model, the Vega PnL under this market scenario would be positive, if we are long this option.

The intuition is that the seasonal model interprets this scenario as saying that even though the December 2014 contract becomes more volatile, it doesn't say anything about the December 2015 contract. On the other hand, the non-seasonal model interprets the market scenario as saying that not only December 2014 contract becomes more volatile, but December 2015 contract also becomes more volatile, during the period from now to the expiration of the 2014 contract.

How can the implied volatility of the December 2015 contract not change? It must be that the volatility of that contract during the period from the expiration of 2014 contract to its own expiration has decreased, to balance the increase of the volatility in its earlier period so that the average volatility for the whole lifetime is still the same.

Which interpretation is correct? Both satisfy the requirement of matching the model to the market implied volatilities for vanilla options. But both are only approximations to what's happening in the real world, and neither is absolutely correct under all circumstances. The truth often lies in between. So, we're interested in seeking a hybrid model that combines the seasonal and non-seasonal calibration strategy. Equipped with this extra flexibility, we will be more capable of capturing the responsive behavior of the market swaption prices, when the term structure of implied volatilities changes.

Below are details on how to calibrate a hybrid model. Let's assume that the volatility ratio is given by the constant vector $p_i = q_i$, and we use the symbol $\varepsilon$ to denote the degree to which we want the hybrid model to be more like the seasonal or non-seasonal model. This "non-fungibility" parameter is a number between 0 and 1, and the hybrid model reduces to the seasonal model when $\varepsilon = 1$ and it becomes the non-seasonal model when $\varepsilon = 0$. Let's also denote the implied volatility, the option expiration and futures expiration of the M'th contract as $\sigma_M, t_M, T_M$ respectively.

The first step is to calibrate a fully seasonal model, and then assign the result "partially" to the T-dependent local volatility function q(T). For this purpose, we recall that in a seasonal model, the scalar $\lambda$ satisfies the equation

$$\sigma_M^2 t_M = \sum_{i,j=1}^{N} \int_0^{t_M} q_i(T_M) q_j(T_M)\, e^{-(\beta_i+\beta_j)(T_M-s)} \rho_{i,j} ds$$

$$= \lambda^2(T_M) \sum_{i,j=1}^{N} q_i q_j \rho_{i,j} e^{-(\beta_i+\beta_j)T_M} \frac{e^{(\beta_i+\beta_j)t_M} - 1}{\beta_i+\beta_j}$$

So

$$\lambda(T_M) = \sigma_M \sqrt{\frac{t_M}{\sum_{i,j=1}^{N} q_i q_j \rho_{i,j} e^{-(\beta_i+\beta_j)T_M} \frac{e^{(\beta_i+\beta_j)t_M} - 1}{\beta_i+\beta_j}}}$$

Now we set the T-dependent local volatility to be a fraction (in the log space) of the solution above, according to the non-fungibility parameter:

$$q_i(T) = q_i^{\varepsilon} \lambda^{\varepsilon}(T)$$

The "residual" fungible implied volatilities $\sigma_M / \lambda^{\varepsilon}(T_M)$ are then to be matched using a boot-strapping approach., the same way as in the non-seasonal model. Recall that the equations to solve for are

$$\sigma_M^2 t_M / \lambda^{2\varepsilon}(T_M) = \sum_{i,j=1}^{N} \int_0^{t_M} p_i(s) p_j(s)\, e^{-(\beta_i+\beta_j)(T_M-s)} \rho_{i,j} ds$$

$$= \sum_{i,j=1}^{N} p_i p_j \rho_{i,j} e^{-(\beta_i+\beta_j)T_M} \sum_{k=1}^{M} \alpha_k^2 \frac{e^{(\beta_i+\beta_j)t_k} - e^{(\beta_i+\beta_j)t_{k-1}}}{\beta_i+\beta_j}$$

We set the t-dependent local volatility to be

$$p_i(t) = p_i^{1-\varepsilon} \alpha(T)$$

It is easy to verify that such calibrated model always matches the implied volatilities of all vanilla options, i.e.

$$\sigma_M^2 t_M = \sum_{i,j=1}^{N} \int_0^{t_M} p_i(s) p_j(s)\, q_i(T_M) q_j(T_M) e^{-(\beta_i+\beta_j)(T_M-s)} \rho_{i,j} ds$$

When $\varepsilon = 0$, the model is equivalent to the fully non-seasonal model, and when $\varepsilon = 1$, the model is equivalent to the fully seasonal model. For the early expiration option example earlier, when $\varepsilon = 1/2$, we will have about half of the vega attributed to December 2014 contract and half to the December 2015 contract.

Finally, how do we calibrate the non-fungibility parameter $\varepsilon$ in the model? Since the main impact of the parameter is to the Vega attribution, we can monitor market swaption prices over a period of time where the term structure of implied volatilities goes through significant changes, and choose $\varepsilon$ such that the responsive behavior of the model matches the market prices the best. This can be tricky though since there could be other factors affecting the changes in swaption prices, especially over an extended period of time.

As an illustration, we pick four month-end dates in 2013, and calculate the IPV (difference between model price and Totem consensus) for our existing WTI swaption positions, using 3 different models: fully non-seasonal, fully seasonal and hybrid with 50% non-fungibility parameter. The model parameters are chosen such that the swaption NPV as of October 8, 2013 are about the same for the 3 models. One can see that for this 4-month period, the hybrid model is more stable than the non-seasonal model, but it is hard to judge between the seasonal and hybrid models. But for oil assets, the hybrid model is naturally preferred over the fully seasonal model.

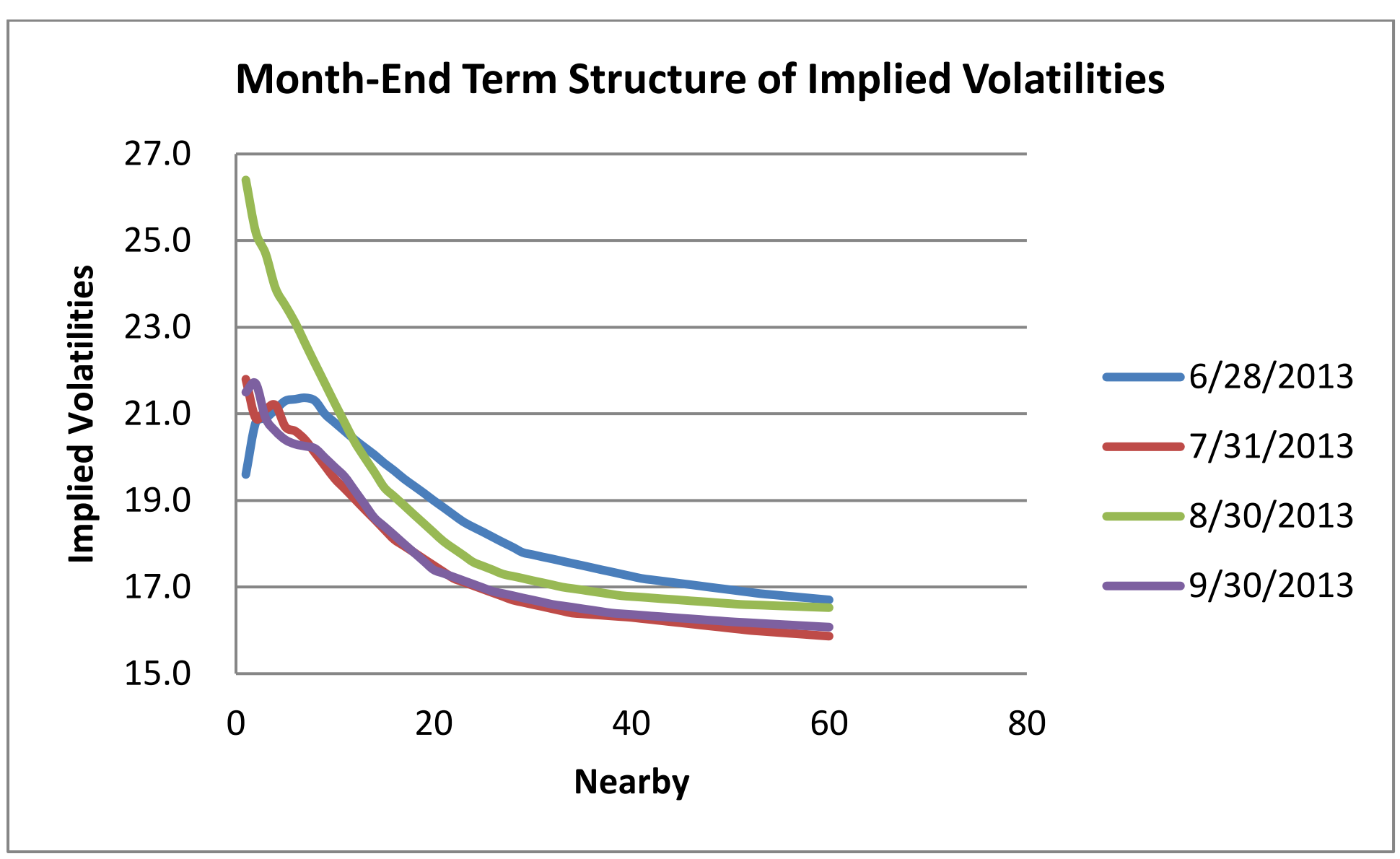

Figure 5. Month-end term structure of implied volatilities.

| Model Params | Non-Seasonal | Seasonal | Hybrid |
| --- | --- | --- | --- |
| Mean Reversion | 0.59 | 0.53 | 0.56 |
| Volatility Ratio | 1.32 | 1.67 | 1.44 |
| Correlation | 0% | 0% | 0% |
| Non-Fungibility | 0% | 100% | 50% |

Table 4. Calibrated model parameters for seasonal, non-seasonal, or hybrid assumptions

| WTI Swaption IPV | Non-Seasonal | Seasonal | Hybrid |
| --- | --- | --- | --- |
| 6/28/2013 | $ 432,064 | $ 504,492 | $ 458,317 |
| 7/31/2013 | $ 444,942 | $ 695,192 | $ 561,262 |
| 8/30/2013 | $ (25,346) | $ 635,440 | $ 300,675 |
| 9/30/2013 | $ 346,722 | $ 555,789 | $ 442,195 |

Table 5. Valuations results under seasonal, non-seasonal or hybrid assumptions

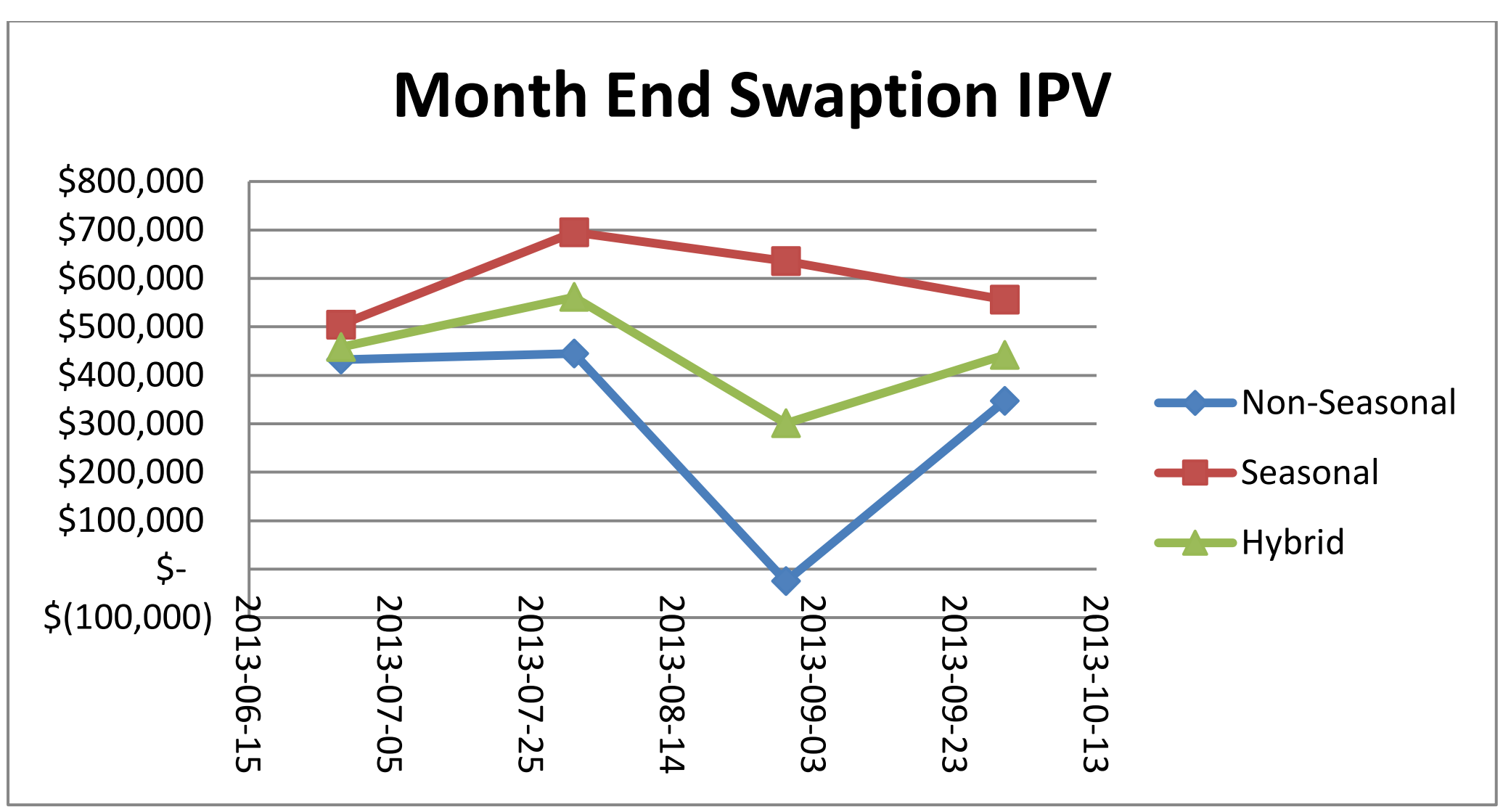

Figure 6. The plots of month-end swaption IPVs

The "curve" risk associated with only the swaptions are exotic in nature and cannot be hedged with vanilla options. The "curve" risk means the uncertainly with how the front and back end of the futures curve tend to move together. If the front end and back end moves in tandem, swaptions should increase in value relative to vanilla options. If the front and the back move in opposite directions, swaptions should decrease in value relative to vanilla options.

There's no way to offset this curve risk effectively with vanilla options. Of course, the most effective way to "hedge" this is to sell the swaptions we bought. But it's quite often that we cannot do this without incurring big costs. Another lesser effective way is to use time spread options as a hedge. But time spread options are as illiquid as swaptions.

The manifestation of this in the model is that the risks to the model parameters, including volatility ratio, mean reversion and correlations are, to the most part, un-hedgeable (unless one sells the swaptions he bought). These parameters are not directly observable from the market either. So how to mark these parameters and manage the uncertainty with these parameters presents a challenge to the model. But there are at least 3 types of mitigation to this challenge.

First, MR imposes a hard limit on how much swaption Vega traders can warehouse. This limits our exposure to the un-hedgeable swaption risks.

Second, VCG verifies the model valuation at least monthly through the IPV process. This ensures that we mark the model to the market expectation of how the "curve" dynamics will actualize.

Finally, we review the historical analysis periodically to ensure we are not stepping into an overheated market. But typically mark-to-market is the rule rather than exception. No amount of historical analysis can tell us for sure what is going to happen in the future.

## 8. Conclusions

In this article, we present a generic model to capture the dynamics of commodity prices. The model describes commodity price as a sum of mean reverting factors. Factors with low mean reversion rate are called Slow/Long Factors that represent uncertainties on long-term fundamentals. Factors with high mean reversion rate are called Fast/Short Factors, and are used to model short term risk factors such as weather, middle-east unrest, economic slowdown, central-bank policy changes, etc. This model can be used to price, hedge, and risk-manage commodity derivatives.

We also propose a generic calibration procedure for the multi-factor model. The calibration procedure consists of an offline step where the mean reversion rates, the ratio of the long and short factor volatilities and the correlation between the long and short factors are determined via historical analysis. This offline step is performed relatively infrequently. There's also an online step of the calibration which happens every time the model is used to price an option or to simulate price paths.

The online calibration ensures that the model prices for vanilla options match exactly the market prices. Finally, there is an introspective step where trading desk adjusts the factor parameters in the offline step to better match the market prices for non-vanilla options such as swaptions and term-settled options. This introspective step happens whenever we have an observation of the market price for these illiquid options.

Empirical study shows the essence of the multi-factor driving of commodity prices. It also shows the model is able to not only value illiquid commodity derivatives accurately but also hedge commodity sensitivities effectively with liquid vanilla options and futures.